\documentclass[11pt]{article}
\usepackage{srcltx}
\usepackage{amsmath}
\usepackage{amssymb}
\usepackage{amsfonts}
\usepackage{latexsym}
\usepackage{textcomp}
\usepackage{appendix}
\usepackage{multirow}
\usepackage{booktabs}
\usepackage{url}
\usepackage{array}
\usepackage{marvosym}
\usepackage[ansinew]{inputenc}
\usepackage{graphics}
\usepackage{graphicx}

\title{LIBOR troubles: anomalous movements detection based on Maximum Entropy}

\author{Aurelio Fern\'andez Bariviera   \\ \footnotesize{Department of Business, Universitat Rovira i Virgili, Av. Universitat 1, 43204 Reus, Spain} \\ \footnotesize{\ttfamily aurelio.fernandez@urv.net} \and M. T. Mart\'in \\ \footnotesize{IFLP-CONICET-UNLP, C. C. 727, 19000 La Plata, Argentina} \and A. Plastino \\ \footnotesize{IFLP-CONICET-UNLP, C. C. 727, 19000 La Plata, Argentina} \and V. Vampa \\ \footnotesize{Math Dept., Engineering Faculty, UNLP} }

\begin{document}
\maketitle

\begin{abstract}

According to the definition of the London Interbank Offered Rate
(LIBOR), contributing banks should give fair estimates of their
own borrowing costs in the interbank market. Between 2007 and
2009, several banks made inappropriate submissions of LIBOR,
sometimes motivated by profit-seeking from their trading
positions. In 2012, several newspapers' articles began to cast
doubt on LIBOR integrity, leading surveillance authorities to
conduct investigations on banks' behavior. Such procedures
resulted in severe fines imposed to involved banks, who recognized
their financial inappropriate conduct. In this paper, we uncover
such unfair behavior by using a forecasting method based on the
Maximum Entropy principle. Our results are robust against changes
in parameter settings and could be of great help for market
surveillance. 


\end{abstract}

\section{Introduction}\label{sec:intro}

London Interbank Offered Rate (LIBOR) was established in 1986 by
the British Banking Association (BBA), who defines LIBOR as ``the
rate at which an individual Contributor Panel bank could borrow
funds, were it to do so by asking for and then accepting
inter-bank offers in reasonable market size, just prior to 11:00
[a.m.] London time''. Every London business day each bank within
the Contributor Panel (selected banks from BBA) makes a blind
submission (each banker does not know what the quotes of the other
Banks are) and a compiler (Thomson Reuters) averages the second
and third quartiles. In other words, LIBOR is the trimmed average
of the expected borrowing rates of leading banks. LIBOR rates are
published for several maturities and currencies.

Over the time LIBOR became a fundamental  interest rate with three
main characteristics: (i) it was viewed as an (intended) measure
of the borrowing cost in the interbank market, (ii) before the
financial crisis, it was interpreted as a risk free rate and (iii)
it is a signal of global credit market conditions. Libor is
enormously influential due to its use for the valuation of
financial products worth trillions of dollars (\cite{BISstat})

The way in which LIBOR is fixed is peculiar, because it does not
arise from actual transactions. It is not the result of the competing forces of supply and demand. There is a panel of banks selected
by the BBA. Each of them should submit their best estimate
according to the following question: ``At what rate could you
borrow funds, were you to do so by asking for and then accepting
inter-bank offers in a reasonable market size just prior to 11
am?'' (\cite{BBATrent}). At some point, individual bank LIBOR
submissions are often regarded as a proxy for the financial health
of the submitting entity. Usually, an employee or group of
employees responsible for cash management in a bank are in charge
of  the daily submission to BBA. They should base their submission
on the money market conditions for the bank, and should not be
influenced by other bank divisions such as the derivatives trading
desks. A fair Libor could signals the state of the interbank money market, and the central banks could act to alleviate frictions in it.

Until May 29, 2008 LIBOR was presumed a pretty honest estimation
of the borrowing costs of prime banks. On that day,
\cite{MollenkampWhitehouse} published an article on the Wall
Street Journal casting doubts on the transparency of LIBOR's
setting, implying that published rates were lower than those
implied by credit default swaps (CDS). Investigations conducted by
several market authorities such as US Department of Justice, the
European Commission, and the Financial Services Authority
(FSA)\footnote{It is noteworthy that the Financial Services Act
2012 renamed FSA as Financial Conduct Authority (FCA), raising the
importance of ``fair conduct'' in financial markets.} detected
data manipulation and imposed severe fines to banks involved in
such illegal procedure.

Several leading banks applied for leniency. Jurists use to say
\textsl{``confessio est probatio probatissima''}, i.e. confession
is the best proof. Therefore, we can accept that, at least, there
was some kind of unfair individual submissions or even worse, a
collusion attempt by a cartel of banks. This manipulation had two
main objectives. On the one hand, low submissions were intended to
give the market a signal of the  bank's own good financial health.
If a bank steadily submits greater rates, this could indicate
problems in raising money, generating concerns regarding  a
underlying solvency problem. On the other hand, some low
submissions could be aimed to earn money from certain portfolio
positions, whose assets are valued according to LIBOR.

The effect of erroneous LIBOR extends beyond the financial
markets. In addition to provide a biased interbank lending cost,
\cite{Stenfors} affirms that  it corrupts a  ``key variable in the
first stage of the monetary transmission mechanism''.

The importance of a good pricing system is based  on its
usefulness for making decisions. As Hayek \cite{Hayek45} affirmed ``we
must look at the price system as such a mechanism for
communicating information if we want to understand its real
function''. If the price system is contaminated, but perceived as
pure, the effect could reach also the real economy, making it
difficult to find a way out the financial crisis.

This rate-rigging scandal made economists to examine  the
evolution of LIBOR rates and compare it with other market rates.
\cite{TaylorWilliams2009} documented the decoupling of the LIBOR
rate from other market rates such as the Overnight Interest Swap
(OIS), Effective Federal Fund (EFF),  Certificate of Deposits
(CDs), Credit Default Swaps (CDS), and Repo rates. They
hypothesize that the reasons for the divergent behavior were due
to expectations of future interest rates and in the accompanying
counterpart risk. \cite{SniderYoule} study individual quotes in
the LIBOR bank panel and corroborate the claim by
\cite{MollenkampWhitehouse} that LIBOR quotes in the US are not
strongly related to other bank borrowing cost proxies. In their
model, the incentive for misreporting borrowing costs is profiting
from a portfolio position. Consequently, the misreporting could
point upwards in one currency and downwards in another one,
depending on the portfolio exposition. The evidence of such
behavior is detected with the formation of a compact cluster of
the different panel bank quotes around a given point.
\cite{AbrantesMetz2011} track daily LIBOR rates over the period
1987 to 2008.

\vskip 3mm

In particular, this paper analyzes the empirical distribution of
the Second Digits (SDs) of the Libor interest rate, and compares
them with the uniform and Benford's distributions. Taking into
account the whole period, the null hypothesis that the empirical
distribution follows either the uniform or the Benford's
distribution cannot be rejected. However, if only the period after
the sub-prime crisis is taken into account, the null hypothesis is
rejected. This result puts into question the ``aseptic'' setting
of LIBOR. In a recent paper  Bariviera \textit{et al.} \cite{Bariviera2015} the authors
uncover strange changes in the information endowment of LIBOR time
series, as measured by two information theory quantifiers, namely
permutation entropy and permutation statistical complexity. Their
results allow to infer some degree of manipulation or, at least,
changes in the underlying stochastic process that governs interest
rate's time series.

Antitrust law enforcement is complex,  because manipulation and fraud can be elegantly camouflaged. An
statistical procedure could  hardly be accepted as legal proof in
a court of law. However, its use by surveillance authorities makes
the attempted manipulation more costly and more difficult to
be maintained. Consequently, we view our proposal as a market watch
mechanism that could make manipulation and/or collusion attempts
more difficult in the future. Additionally, an efficient
overseeing mechanism could increase the incentives to apply for
leniency at earlier stages of the manipulation
(\cite{AbrantesSokol2012}.)

The aim of this paper is to show that a forecasting method based
on Maximum Entropy Principle (MaxEnt) is very useful not only to
produce accurate forecasts, but also to detect some anomalous
situations in time series. In particular, we claim that, in
absence of data manipulation, forecast accuracy should be
approximately the same at all times under examination. On the
contrary, manipulation would produce more  predictable
consequences, increasing the predictive-power of our model, that
we apply here to LIBOR and other UK interest rates.

This paper is  organized as follows. Section \ref{sec:ME}
describes our methodology based on the Maximum Entropy method.
Section \ref{sec:data} describes the data used in the paper and
deals with the results obtained with the proposed methodology.
Finally, section \ref{sec:conclusions} draws the main conclusions.

\section{MaxEnt approach for predictions in time-series}\label{sec:ME}

In a recent paper, Mart\'in \textit{et al. }\cite{Martin2014} developed an information
theory based method  for time series prediction. Given its
outstanding results in approaching the true dynamics underlying a
given time series, we believe that it is a suitable method to
apply here. In order to make the paper self-contained, we review
below the description of the method, taken from \cite{Martin2014}.

\noindent The  behavior of a dynamical  system can be registrated as
a time series i.e.  a sequence of measurements
$\{ v(t_n), n=1,
\ldots , N\}$ of an observable of the system
at discrete times $t_n$, where
$N$ is the length of the time series.

\noindent The Takens theorem of 1981 asserts that, for $T \in
\mathbb{R}$, $T>0$, there exists a functional form of the type,
\begin{equation} \label{mapping}
v(t+T)=F({\bf v}(t)),
\end{equation}
where
\begin{equation} \label{vector v}
{\bf v}(t)=[v_1(t),v_2(t),\ldots,v_d(t)],
\end{equation}
and  $v_i(t)=v(t-(i-1) \Delta)$, for $i=1,\ldots,d$. $\Delta$ is
the time lag and $d$ is the embedding dimension of the
reconstruction.  $T$ represents the \emph{anticipation time} and
it is of fundamental importance for a prediction model.

\noindent We will consider (as in [\cite{Martin2014}) and
references therein] a particular representation for the mapping
function of Eq. (\ref{mapping}), expressing it, using Einstein's
summation notation,  as an expansion of the form

\begin{eqnarray}
F^*({\bf v}(t))=&a_{0}+a_{i_1} v_{i_1}(t)~ +a_{i_1 i_2}v_{i_1}(t)~
v_{i_2}(t)~ + ~a_{i_1 i_2 i_3}v_{i_1}(t)~ v_{i_2}(t)~v_{i_3}(t)~ +
\ldots  \label{Fasterisco}
  \\ \nonumber & +a_{i_1 i_2... i_{n_p}}v_{i_1}(t)~
v_{i_2}(t)~\ldots v_{i_{n_p}}(t)~,
\end{eqnarray}
where $1 \le i_k \le d $ with $1 \le k \le n_p$  and  $n_p$ being
an  adequately chosen polynomial degree so as to to series-expand
the mapping $F^*$. The number of parameters in
Eq.(\ref{Fasterisco})  corresponding to  the terms of  degree  $k$
depends on the embedding dimension and  can be calculated using
combination with repetitions,
\begin{equation} \label{variaciok}
\left(
\begin{array}{c}
  d \\
  k\\
\end{array}%
\right)^* = \frac{(d+k-1)!}{k! (d-1)!}
\end{equation}

\noindent Accordingly, the length of the vector of parameters
$\textbf{a}$, $N_c$, is
\begin{equation} \label{Nc}
N_c=
 \sum_{k=1}^{n_p}\left(%
\begin{array}{c}
  d \\
  k\\
\end{array}%
\right)^*
\end{equation}

\noindent The computations are  made on the basis of a specific
information supply,  given by $M$ points of the series
\begin{equation}\label{Ecuacion5}
\{{\bf v}(t_n),v(t_n+T)\},~~~n=1,\ldots,M.
\end{equation}

\noindent Given the data set in Eq. (\ref{Ecuacion5}), the
parametric mapping  in Eq. (\ref{Fasterisco}) will be determined
by  the following condition,

\begin{equation}\label{Ecuacion7}
F^*({\bf v}(t_n))=v(t_n+T)~~~~~~n=1,\ldots,M ,
\end{equation}
which can be expressed in matrix form as,
\begin{equation}\label{Ecuacion12}
 W   \textbf{a} =\textbf{v}_T,
\end{equation}
where   $W$ is a matrix of size $ M \times N_c$, whose $n$-th row
is \\ $[1,v_{i_1}(t_n), v_{i_1}(t_n) v_{i_2}(t_n),\ldots ,
v_{i_1}(t_n) v_{i_2}(t_n) \ldots v_{i_{n_p}}(t_n)] $ (Cf.
Eq.(\ref{Fasterisco})) and \\ $(\textbf{v}_T)_n=v(t_n+T)$, for
$n=1,\ldots,M$.   Shannon's entropy, defined for a discrete random
variable, can be extended to situations for which the random
variable under consideration is continuous.

\noindent In order to infer coefficients which are consistent with
the data we shall assume that each set $\textbf{a}$ is realized
with probability $P(\textbf{a})$. Thus,
 a normalized probability distribution over the possible sets $\textbf{a}$ is introduced,
\begin{equation}\label{EcuacionP}
 \int_{I} P(\textbf{a}) \  d\textbf{a} =1 ,
\end{equation}
where $d\textbf{a}=da_1da_2 \cdots da_ {N_c}$ and $N_c$ is the
number of parameters of the model.

\noindent The problem then becomes that of  finding
$P(\textbf{a})$ subject to the requirement that the associated
entropy $H$  be maximized, since this is the best way of avoiding
any bias. The expectation value of $\textbf{a}$, is defined by
\begin{equation}\label{EcuacionEa}
 \left\langle \textbf{a} \right\rangle = \int_{I}{P(\textbf{a}) \textbf{a}\  d\textbf{a} }.
\end{equation}
\noindent Consider the  continuous random variable $\textbf{a}$
with probability density function $ p(\textbf{a})$ on $I$ and $\,
I=(-\infty,\infty)$. The entropy is given by

\begin{equation}\label{EcuacionH}
\displaystyle H(\textbf{a})=-\int_{I}{P(\textbf{a})\ \ln\
P(\textbf{a})\ d\textbf{a}},
\end{equation}
whenever it exists, and the  relative entropy reads

\begin{equation}\label{EcuacionHr}
\displaystyle H=-\int_{I} P(\textbf{a})\ \ln\ \frac
{P(\textbf{a})} {P_0(\textbf{a})} \ d\textbf{a},
\end{equation}
where $P_0(\textbf{a})$ is an appropriately chosen a priori
distribution.

\noindent This measure exhibits many of the properties of a
discrete entropy but, unlike the entropy of a discrete random
variable, that for a continuous random variable may be infinitely
large, negative, or positive (Ash, 1965 [6]). \noindent We
characterize, via the entropic maximum principle, various
probability distributions, subject to the constraints
Eq.(\ref{EcuacionP}) and Eq.(\ref{Ecuacion12}) for the expectation
$\left\langle \textbf{a} \right\rangle$ of $\textbf{a}$. \noindent
The method for solving this constrained optimization problem is to
use Lagrange multipliers for each of the operating constraints and
maximize the following functional with respect to $P(\textbf{a})$,

\begin{equation}\label{EcuacionHrmultip}
 J=- \left[\int_{I} P(\textbf{a}) \ln \frac {P(\textbf{a})} {P_0(\textbf{a})} d\textbf{a} + \lambda_0 \left[\int_{I} P(\textbf{a}) d\textbf{a}-1\right] +  {\lambda}^t \int_{I} \left[W P(\textbf{a}) \textbf{a} - \textbf{v}_T \right]  d\textbf{a} \right],
\end{equation}
where $\lambda_0$  and $\lambda$  are Lagrange multipliers
associated, respectively, with the normalization condition and
with the constraints, Eq.(\ref{EcuacionP}) and
Eq.(\ref{Ecuacion12}).

\noindent Taking the functional derivative with respect to
${P(\textbf{a})}$  we get
\begin{equation}\label{Ecuacion20}
\frac{ \partial J}{\partial P(\textbf{a})} = \ln
\left(\frac{P(\textbf{a})}{P_0(\textbf{a})}\right) +1 + \lambda_0
+ {\lambda}^t  W \textbf{a} =0,
\end{equation}
which implies that the maximum entropy distribution must have the
form
\begin{equation}\label{Ecuacion25}
 P(\textbf{a})= \exp -(1 +  \lambda_0 )  \exp(\lambda^t W \textbf{a}) P_0(\textbf{a})
\end{equation}

\noindent If the a priori probability distribution
$P_0(\textbf{a})$  is chosen to be proportional to $\exp(
-\frac{1}{2} \textbf{a}^t [\sigma^2]^{-1} \textbf{a})$,  where
$\sigma^2$ is the covariance matrix, a Gaussian form  for the
probability distribution $P(\textbf{a})$ is obtained, with
\begin{equation}\label{Ecuacion30}
\left\langle \textbf{a} \right\rangle = -\sigma W^t \lambda
\end{equation}

\noindent Considering  Eq.(\ref{Ecuacion12}), the Lagrange
multipliers $\lambda$ can be eliminated:

\begin{equation}\label{Ecuacion40}
\lambda= -\sigma^{-1} (W W^t)^{-1} \textbf{v}_T,
\end{equation}
and one can thus write

\begin{equation}\label{Ecuacion45}
\left\langle \textbf{a} \right\rangle = W^t  (W W^t)^{-1}
\textbf{v}_T .
\end{equation}
\noindent The matrix $ W^t  (W W^t)^{-1}$ is  known as the
Moore-Penrose pseudo-inverse of the matrix $W$ (see
\cite{Martin2014} and references therein). Consequently,
 this result shows that the  maximum entropy principle  coincides with a least square criterion.
\noindent Once the pertinent parameter vector  $\textbf{a}$ is determined, it is
used to predict  {\bf new} series' values, $\widehat v(t_n+T))_{n=1,\ldots, M_P}$, according to

\begin{equation} \label{Prediccion}
(\widehat v(t_n+T))_{n=1,\ldots, M_P}= \widehat{ W} \textbf{a},
\end{equation}
 where $\widehat W$ is the matrix of size $ M_P \times N_c $ (see Eq.(\ref{Ecuacion12})), obtained using  $\widehat v(t_n)$ values.

\section{Data and results }\label{sec:data}

We analyze the Libor in pound Sterling. The data span is from 01/01/1999 until  21/10/2008, with a total of 2560 datapoints. All data were retrieved from DataStream.

In this section we present the results obtained using the  methodology  proposed in Section \ref{sec:ME}. We consider the embedding dimension $d=4$ and the polynomial degree $n_p=2$. The length of the vector of parameters, according to Eq. \ref{Nc} is $N_c=15$.

We fit our model with $M=700$ datapoints, corresponding to approximately two and a half years beginning on  01/01/1999. Once the model's parameters were determined, we forecasted the rest of the time series, up to 21/10/2008.

\begin{figure}[!ht]
\center
\includegraphics[width=16cm,height=10cm]{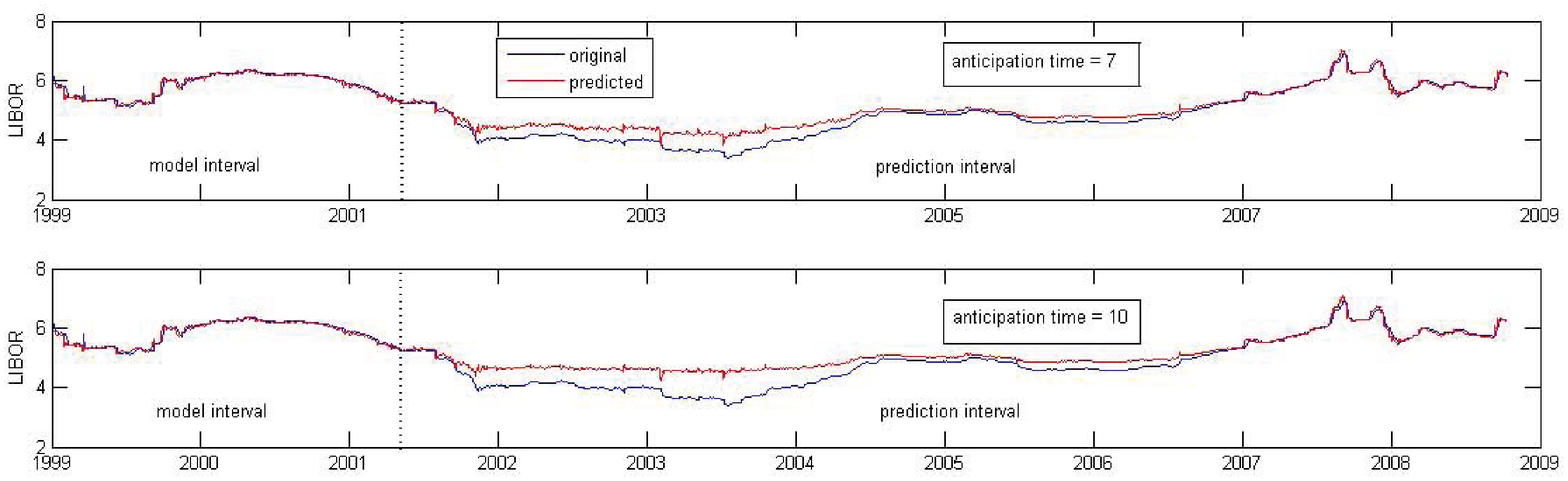}
\caption{Original and forecasted time series for different anticipation times}
\label{fig:T1}
\end{figure}

\begin{figure}[!ht]
\center
\includegraphics[width=16cm,height=10cm]{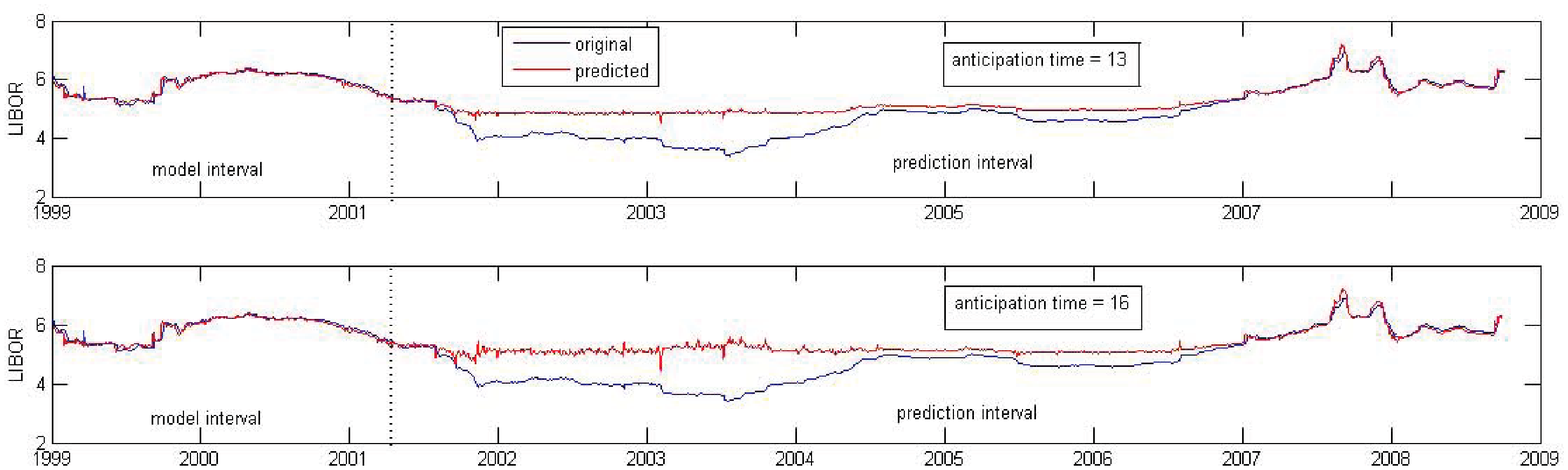}
\caption{Original and forecasted time series for different anticipation times}
\label{fig:T2}
\end{figure}

In the figures \ref{fig:T1} and \ref{fig:T2}  the original time series values and the predicted ones are overlapped (blue and red refer to original and
predicted values, respectively) for different anticipation time values.  The time-interval between the beginning of the time series and the vertical dashed lines corresponds to the model interval, used to estimate the parameters. The other part corresponds to the out-of-sample forecasts.

In order to prove the robustness of our proposal we did forecast for different anticipation times (T=\{7, 10, 13, 16\} days).

\begin{figure}[!ht]
\center
\includegraphics[scale=.4]{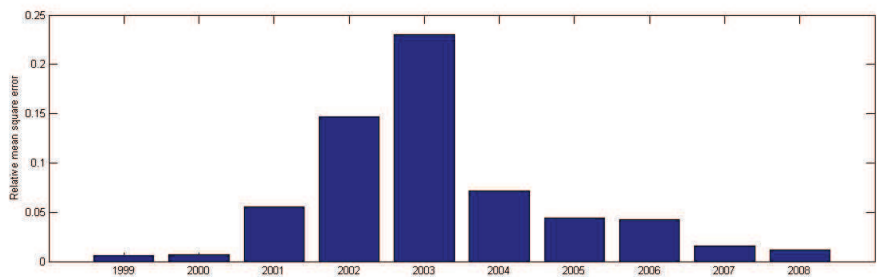}
\caption{Relative mean square errors }
\label{errorpor}
\end{figure}

We can observe in figures \ref{fig:T1} and \ref{fig:T2} that, as
expected, during the model interval period, the original and the
predicted time series are very close. This is the consequence of
the adequate fitting power of the model. As is the case for any
forecast method, one tries to mimic the behavior of the time
series to be estimated. When we move into the (out of the sample)
prediction interval, we note that during the first months, our
method behaves very well. We expect that, as  economic theory
affirms, competitive prices should behave randomly
(\cite{Samuelson65}). Consequently, if we assume that the time
series under study is generated by a memoryless stochastic
process, accurate forecasts are not possible. In spite of the fact
that the original time series changes, we can see that the
predicted time series is rather constant between 2002 and 2007.
This is the consequence of the stochastic nature of the original
time series. The prediction performance is very poor. In addition,
the distance between the original and the predicted series in this
period increases monotonically with the anticipation time, as
expected. Surprisingly enough, beginning with 2007, our model
begins to fit real data very well. Predicted time series moves
\textit{pari passu} with the original one, even during the large
increases during 2008. A similar analysis can be done with
reference to figure \ref{errorpor}. In that figure, we display the
relative mean square error between the original and forecasted
time series, year by year.

What could make  the same model to change its forecast accuracy in
such  dramatic fashion? According to Wold's theorem
(\cite{Wold1954}), a time series can be separated into a
deterministic part and an stochastic part. If the stochastic part
dominates the behavior of the time series, forecast is
unsuccessful. This is what we can observe between 2002 and the end
of 2006. On the contrary, beginning in 2007, and until the end of 2008,
prediction becomes very accurate. Given that the prediction model
is the same for both periods, we conjecture that the time series
is dominated by a deterministic process in the last of the two
periods. Recalling the literature review of Section
\ref{sec:intro}, we can state that this result is an indirect
proof of LIBOR manipulation. We emphasize that such
``manipulation'' necessarily comprises the contamination of the
time series with a deterministic device, which was detected by the
MaxEnt model.

\section{Conclusions \label{sec:conclusions}}

In this paper we present a novel  prediction method based on the
MaxEnt principle. Taking into account its previous performance
(\cite{Martin2014}), we believe it is suitable for the study of the
``Libor Case''. We study Libor time series between 1999 until
2009. Based on the prediction accuracy of our method, we are able
to detect two distinctive regimes. The first one, extends between
2002 and the end of 2006. In this period the time series behaves
as predicted by standard economic theory, reflecting the random
character of prices in competitive environments. The prediction
power is, consequently, poor. The second time-period spans  2007 and
2008. In this period the time series changes its
regime, moving to a more predictable one. We can safely think that
a deterministic device was introduce into the Libor setting. This
situation takes place at the time that what was called by the
newspapers as the ``Libor manipulation'' one.  As a consequence,
our paper is able to detect such manipulation, using exclusively
data from Libor time series. We would like to emphasize the
relevance of advanced statistical models in market's watch
mechanisms. Our results could be of interest to surveillance
authorities, given the importance of fair market conditions in
free market economies.

\bibliographystyle{plain}
\bibliography{liborbib}

\end{document}